\documentclass[10pt]{report}
\usepackage{graphicx}
\usepackage{amssymb}
\usepackage{color}
\usepackage{makeidx}
\usepackage[numbers,sort&compress]{natbib}
\bibpunct{[}{]}{,}{n}{}{,}

\renewcommand{\abstract}[1]{{ \footnotesize \noindent {\bf Abstract} #1 \\}}
\renewcommand{\author}[1]{\subsubsection*{\it#1}}
\newcommand{\address}[1]{\subsubsection*{\it#1}}

\begin{document}

\chapter{Cometary diversity and cometary families}

%\title{Cometary diversity and cometary families}

%\maketitle

\author{Jacques Crovisier}

\address{Observatoire de Paris \\
5, place Jules Janssen, F-92195 Meudon, France\\
email: \texttt{jacques.crovisier@obspm.fr}}

%\date{\today}

%\maketitle

\abstract{Comets are classified from their orbital characteristics into two
separate classes: nearly-isotropic, mainly long-period comets and ecliptic,
short-period comets.  Members from the former class are coming from the Oort
cloud.  Those of the latter class were first believed to have migrated from the
Kuiper belt where they could have been accreted in situ, but recent orbital
evolution simulations showed that they rather come from the trans-Neptunian
scattered disc.  These two reservoirs are not where the comets formed: they were
expelled from the inner Solar System following interaction with the giant
planets.  If comets formed at different places in the Solar System, one would
expect they show different chemical and physical properties.  In the present
paper, I review which differences are effectively observed: chemical and
isotopic compositions, spin temperatures, dust particle properties, nucleus
properties...  and investigate whether these differences are correlated with the
different dynamical classes.  The difficulty of such a study is that
long-period, nearly-isotropic comets from the Oort cloud are better known, from
Earth-based observations, than the weak nearly-isotropic, short-period comets.
On the other hand, only the latter are easily accessed by space missions.}

\section{Introduction}

\begin{figure}
\centering
\includegraphics[width=\hsize]{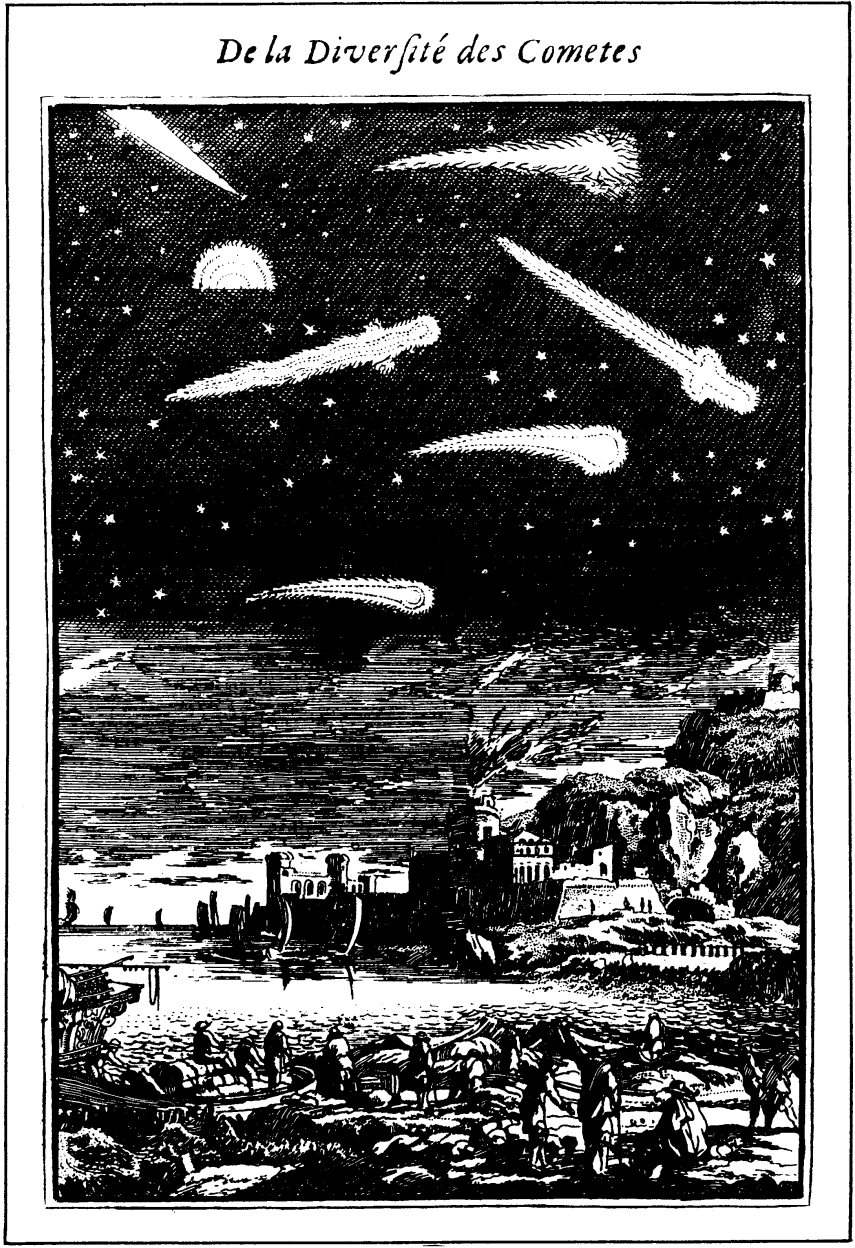}
\caption{Cometary diversity, from a 17th century engraving.  The exact origin is
unknown, but a similar plate was published under a different title in
``Description de l'Univers" (Paris, 1683) by Allain Manesson Mallet
(1630--1706).  At that time, the diversity of comets was remarked from their
appearance.}
\label{fig:diversite}
\end{figure}

There are not two comets alike.  These objects show an extraordinary diversity
(Fig.~\ref{fig:diversite}).  Besides the multiplicity of their appearance, the
diversity of comets is twofold:

\begin{itemize}
    
    \item diversity of orbits, from which different dynamical classes of comets
    have been clearly defined;
    
    \item physical and chemical diversity.  Several attempts to base a taxonomy
    of comets upon their chemical diversity were made.
    
\end{itemize}

Our present knowledge of the nature of comets, of their formation and evolution,
is reviewed in the chapters of \citep{fest+:2005}.  Two fundamental questions
arise, related to the diversity of comets and to the history of their formation:

\begin{itemize}
    
    \item are the different dynamical classes of comets related to different
    formation scenarios, and more specifically, to different sites of formation?
    
    \item is the physical/chemical diversity of comets due to different
    formation conditions?  Is it linked to the dynamical differences?
    
\end{itemize}

\begin{figure}
\centering
\includegraphics[width=0.7\hsize,angle=270.]{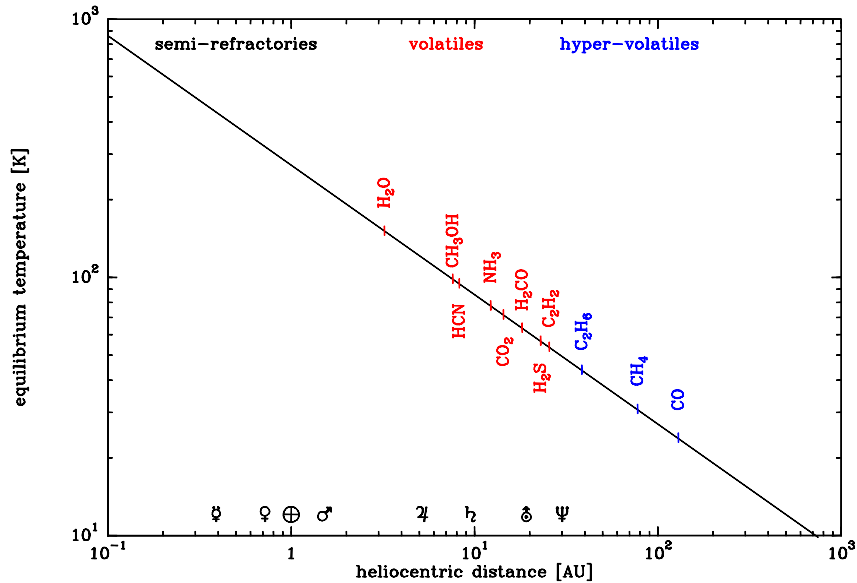}
\caption{The snow lines for various cometary volatiles.  The
sublimation/condensation temperatures of \citep{yama:1985} and a temperature
profile $T = 273\, r_h^{-0.5}$ are assumed.  It must be noted that this law and
the place of the planets, as represented in the figure, pertain to the present
Solar System and may greatly differ when comets formed in the early Solar
System.}
\label{fig:sublime}
\index{molecule!H$_2$O}
\index{molecule!HCN}
\index{molecule!CO}
\index{molecule!CO$_2$}
\index{molecule!CH$_3$OH}
\index{molecule!H$_2$CO}
\index{molecule!H$_2$S}
\index{molecule!CH$_4$}
\index{molecule!C$_2$H$_2$}
\index{molecule!C$_2$H$_6$}
\index{molecule!NH$_3$}
\index{molecule!C$_2$H$_4$}
\index{sublimation temperature}
\index{snow line}
\index{hypervolatile}
\end{figure}

The ices contained in comet nucleus comprise molecules of very different
volatility \citep{yama:1985}.  One would expect that the spread of
sublimation/condensation temperatures of these molecules leads to a correlation
between formation sites and composition, and that the \textit{snow lines} for
the different species follow the temperature profile of the Solar System (i.e.,
the equilibrium temperature as a function of heliocentric distance $r_h$).  The
most volatile molecules (the so-called \textit{hypervolatiles}) such as CO$_2$
and CH$_4$ can only condense only far from the Sun ($r_h > 50$~AU) whereas water
condenses at $r_h > 3$~AU (Fig.~\ref{fig:sublime}).

\index{cometary ices} \index{snow line} \index{hypervolatile}
\index{astrochemistry} \index{sublimation temperature} 
\index{Solar Nebula}

Chemistry in the protosolar nebula is also expected to be highly $r_h$-dependent
\citep{berg+:2007}.  However, an extensive radial mixing of matter in the
protoplanetary Solar Nebula is now testified by several clues (e.g., the
presence of both amorphous and crystalline silicates).  It would tend to make
the chemical composition uniform within the Nebula.  But it does not preclude a
segregation of the various ice components according to their sublimation
temperatures.

On the other hand, short-period comets, which experienced many returns close to
the Sun, are likely to present significant modifications of their nucleus outer
layers due to heating and sublimation fractionation.

The purpose of the present paper is to examine whether the cometary diversity
could be linked to different origins of these bodies.  The dynamical history of
comets and the possible existence of several reservoirs have already been
extensively discussed in the literature (e.g., in several chapters of
\citep{fest+:2005}).  A summary is given in Section \ref{section:orbits}.  A
comparative presentation of the available data on the nature of comets,
according to their different dynamical classes, is given in Section
\ref{section:observations}.  The possible links between cometary diversity,
cometary reservoirs, and their sites of formation, deserve to be analysed and
discussed in the light of recent results; this is done in Sections
\ref{section:chemistry}, \ref{section:physics} and \ref{section:discus}.

\section{Cometary orbits and cometary reservoirs}
\label{section:orbits}

\index{comets!Jupiter-family comets}
\index{comets!Oort-cloud comets}
\index{comets!Kuiper-belt comets}
\index{comets!Main Belt comets}
\index{comets!Halley-type comet}
\index{comets!ecliptic comets}
\index{comets!nearly-isotropic comets}
\index{Oort cloud}
\index{Kuiper belt}
\index{cometary orbits}
\index{Chiron}
\index{Centaurs}

\subsection{Cometary orbits}

Following the now classical view, there are two main classes of comets according
to their orbital characteristics.  \textit{Nearly-isotropic comets} have
arbitrary orbital inclinations; they comprise long-period comets ($\gtrsim 200$
years) and Halley-type comets of shorter orbital periods.  \textit{Ecliptic
comets} (aka Jupiter-family comets) have short orbital periods (typically
$\lesssim 20$ years) and low orbital inclinations (typically $< 20^\circ$);
their orbital evolution is governed by interaction with Jupiter.

A convenient dynamical indicator, used to separate these two classes, is the
Tisserand parameter: \index{Tisserand parameter}

\begin{equation}
    T_J = \frac{a_J}{a} + 2 \sqrt{\frac{a}{a_J} (1 - e^2)} \cos i,
\end{equation}

\noindent where $a_J$ is the semi-major axis of Jupiter (5.20~AU), $a$ that of
the comet orbit, $e$ its eccentricity and $i$ its inclination.  For
Jupiter-family comets, $2 < T_J < 3$, whereas long-period and Halley-type comets
have $T_J < 2$ and asteroids have $T_J > 3$.

\begin{figure}
\centering
\includegraphics[width=\hsize, bb=145 500 555 725,clip]{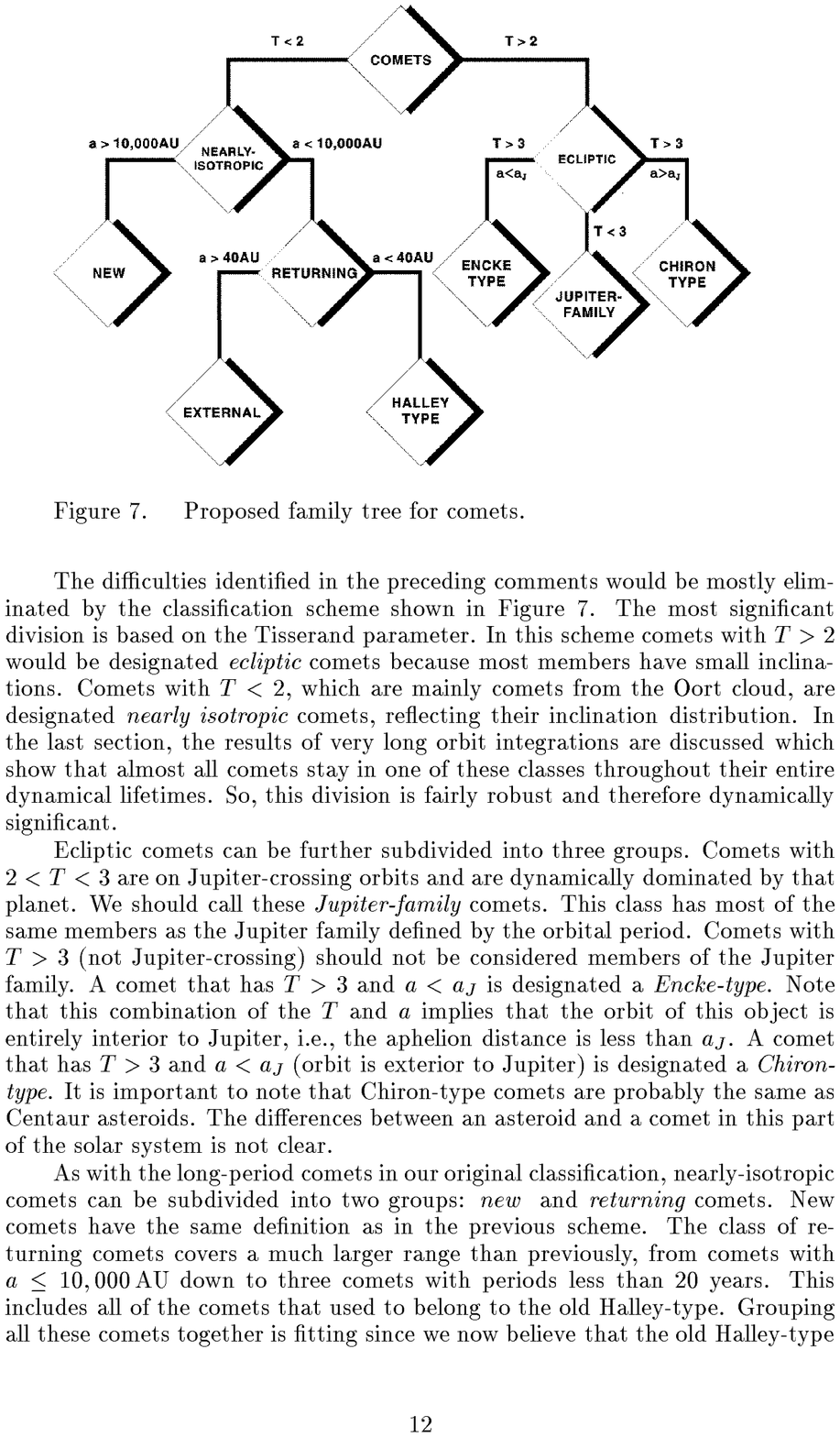}
\caption{Comet families according to Levison \citep{levi:1996}, using
the Tisserand parameter $T$ and the semi-major axis $a$ as criteria.}
\label{fig:levison}
\index{comets!Jupiter-family comets}
\index{comets!Oort-cloud comets}
\index{comets!Halley-type comet}
\index{comets!ecliptic comets}
\index{comets!nearly-isotropic comets}
\index{Chiron}
\index{comet!1P/Halley} \index{comet!2P/Encke} 
\end{figure}

Levison \citep{levi:1996} proposed a slightly modified family tree for comets,
from their Tisserand parameter and their semi-major axis
(Fig.~\ref{fig:levison}).  Chiron-type comets (aka Centaurs --- only a handful
of them are known to be active) and Encke-type comets (only two members are
known to date) are now introduced.

Quite recently, a new class, \textit{Main-belt comets}, was introduced
\citep{hsie-jewi:2006}.  It includes objects with typical orbits of Main Belt
asteroids which show a low level of activity.  It is possible that many objects
presently classified as \textit{asteroids} belong to this class.  Such objects
have rather stable orbits and are probably not dynamically related to the other
classes of comets.  Since there are only three such comets known to date, for
which the chemical and physical characteristics are practically unknown, these
objects will not be considered here.  (Main Belt comets may also be considered
as \textit{activated asteroids} \citep{lica+:2007}.)

\subsection{Where comets are stored}

\index{Oort cloud}
\index{Kuiper belt}
\index{scattered disc}

The origin of comets has been for a long time the subject of debates.  Before
being recognized as Solar System objects, comets were considered as atmospheric
phenomena, or to originate from the Galaxy.  Flammarion and Proctor in the 19th
century, and more recently Crommelin and Vsekhsvyatskii, argued that
Jupiter-family comets were expelled from giant planets \citep{vsek:1964}.

The Oort cloud \cite{oort:1950}, already described by Schiaparelly and \"{O}pik
\cite{schi:1910, opik:1932}, was introduced to explain the continuous input of
new long-period comets.

The belt of trans-Neptunian objects now known as the Kuiper belt
\citep{kuip:1951} was already described by Leonard and Edgeworth
\citep{kuip:1951, leon:1930, edge:1949}.  Following the discovery of the first
batch of trans-Neptunian objects in the nineties, it was readily accepted as the
reservoir of ecliptic comets.  However, the orbits of Kuiper-belt objects appear
to be very stable and unlikely to evolve into short-period comets.  The current
idea is that ecliptic comets rather come from another class of trans-Neptunian
objects, the scattered disc, whose orbits are more eccentric and show
significant inclination \citep{morb:2005, dunc-levi:1997, glad:2005}.  In fact,
the border between the inner part of the Oort cloud and the scattered disc does
not seem to be clearly established.

\subsection{Where comets formed}

Numerical simulations of orbital evolutions can now constrain more precisely the
possible sites of formations of comets \citep{morb:2005, gome+:2005,
dunc-levi:1997, glad:2005} .

Nearly-isotropic comets, formed in the Jupiter--Uranus region, were ejected to
the Oort cloud, then coming back as long-period comets and evolving to
Halley-type comets.

For some time, it was believed that ecliptic comets were formed in the Kuiper
belt beyond Neptune, then migrated to become Jupiter-family comets.  Following
this paradigm, nearly-isotropic comets were dubbed \textit{Oort-cloud comets},
and Jupiter-family comets, \textit{Kuiper-belt comets}.  The two classes of
comets would have then been formed at quite different distances from the Sun:
the ``cold" Kuiper-belt comets in the trans-Neptunian region, and the warmer
Oort-cloud comets in the Jupiter--Uranus region, expected to be comparatively
volatile-depleted.

We now know that ecliptic comets rather come from the scattered disc, and that
they were not formed there, but were expelled from the intra-Neptunian Solar
System, like the Oort-cloud comets.  What are precisely the regions where the
two classes of comets formed, and are there indeed different, is presently a
moot point.  Orbital evolution simulations have shown that the orbits of the
giant planets dramatically changed in the first steps of the Solar System
history, so that their present places may greatly differ from the places they
had when comets formed and were expelled \citep{gome+:2005}.

To complicate the situation, a comet nucleus may result from the accretion of
planetesimals of different origins and different compositions.  This could be
tested by observing comets that fragmented in different components, as did
C/1999 S4 (LINEAR) and 73P/Schwassmann-Wachmann 3.

\index{split comets}
\index{comet!C/1999 S4 (LINEAR)}
\index{comet!73P/Schwassmann-Wachmann 3}

\section{Available observations}
\label{section:observations}

\begin{table}
\caption{Remote sensing observing conditions for a selection of comets}
\label{table:remote}
\begin{center}
\begin{tabular}{llcccc}
\hline
comet & date & $r_h$ & $\Delta$ & $Q$ & $FM$ \\
      &      & [AU]  & [AU]     & [$10^{28}$ s$^{-1}$] \\
\hline
\hline
1P/Halley                    & January 1986   & 0.7 & 1.5  & 120 &  80 \\
C/1986 P1 (Wilson)           & May 1987       & 1.3 & 1.0  &  12 &  12 \\
C/1989 X1 (Austin)           & April 1990     & 1.2 & 0.25 & 2.5 &  10 \\
C/1990 K1 (Levy)             & August 1990    & 1.3 & 0.45 & 25  &  55 \\
109P/Swift-Tuttle            & November 1992  & 1.0 & 1.3  & 50  &  40 \\
C/1996 B2 (Hyakutake)        & March 1996     & 1.1 & 0.11 & 25  & 225 \\
C/1995 O1 (Hale-Bopp)        & April 1997     & 0.9 & 1.4  &1000 & 700 \\
C/1999 S4 (LINEAR)           & July 2000      & 0.77& 0.38 & 10  &  25 \\
C/1999 T1 (McNaught-Hartley) & December 2000  & 1.2 & 1.6  & 10  &   6 \\
C/2001 A2 (LINEAR)           & June 2001      & 1.0 & 0.24 & 10  &  40 \\
C/2000 WM$_1$ (LINEAR)       & December 2001  & 1.2 & 0.32 &  4  &  12 \\
153P/2002 C1 (Ikeya-Zhang)   & April 2002   & 1.0 & 0.40 & 25  &  60 \\
C/2001 Q4 (NEAT)             & May 2004       & 1.0 & 0.32 & 20  &  60 \\
C/2002 T7 (LINEAR)           & May 2004       & 0.8 & 0.27 & 10  &  25 \\
C/2003 K4 (LINEAR)           & December 2004  & 1.4 & 1.2  & 15  &  12 \\
C/2004 Q2 (Machholz)         & January 2005   & 1.2 & 0.35 & 25  &  70 \\
\hline
\hline
22P/Kopff                    & April 1996     & 1.7  & 0.9  & 3.5 & 4 \\
21P/Giacobini-Zinner         & October 1998   & 1.2  & 1.0  & 3   & 3 \\
19P/Borrelly                 & September 2001 & 1.36 & 1.47 & 3   & 2 \\
2P/Encke                     & November 2003  & 1.0  & 0.25 & 0.5 & 2 \\
9P/Tempel~1                  & July 2005      & 1.5  & 0.77 & 1   & 1.3 \\
73P/Schwassmann-Wachmann~3   & May 2006       & 1.0  & 0.08 & 2   & 25  \\
8P/Tuttle                    & January 2008   & 1.1  & 0.25 & 3   & 12 \\
85P/Boethin                  & December 2008  & 1.1  & 0.9  & 3   & 3.3 \\
67P/Churyumov-Gerasimenko    & Mars 2009      & 1.24 & 1.7  & 1   & 0.6 \\
103P/Hartley~2               & October 2010   & 1.1  & 0.12 & 1.2 & 10 \\
45P/Honda-Mrkos-Pajdu\v{s}\'akov\'a & August 2011    & 1.0  & 0.06 & 0.5 &  8 \\
\hline
\end{tabular}
\end{center}

Top part: recent nearly-isotropic comets.\\
Bottom part: recent and future ecliptic comets.\\
Date is for best observing conditions. $Q$ = water production rate at that date.\\
$FM = Q/\Delta$ = Figure of merit.\\
\index{comet!1P/Halley}
\index{comet!2P/Encke}
\index{comet!8P/Tuttle}
\index{comet!9P/Tempel 1}
\index{comet!19P/Borrelly}
\index{comet!21P/Giacobini-Zinner}
\index{comet!22P/Kopff}
\index{comet!45P/Honda-Mrkos-Pajdu\v{s}\'akov\'a}
\index{comet!67P/Churyumov-Gerasimenko}
\index{comet!73P/Schwassmann-Wachmann 3}
\index{comet!85P/Boethin}
\index{comet!103P/Hartley 2}
\index{comet!109P/Swift-Tuttle}
\index{comet!122P/de Vico}
\index{comet!153P/2002 C1 (Ikeya-Zhang)}
\index{comet!C/1986 P1 (Wilson)}
\index{comet!C/1989 X1 (Austin)}
\index{comet!C/1990 K1 (Levy)}
\index{comet!C/1995 O1 (Hale-Bopp)}
\index{comet!C/1996 B2 (Hyakutake)}
\index{comet!C/1999 S4 (LINEAR)}
\index{comet!C/1999 T1 (McNaught-Hartley)}
\index{comet!C/2000 WM$_1$ (LINEAR)}
\index{comet!C/2001 A2 (LINEAR)}
\index{comet!C/2001 Q4 (NEAT)}
\index{comet!C/2002 T7 (LINEAR)}
\index{comet!C/2003 K4 (LINEAR)}
\index{comet!C/2004 Q2 (Machholz)}
\end{table}

\begin{table}
\caption{Space exploration of comets}
\label{table:space}
\begin{center}
\begin{tabular}{llccccc}
\hline
Comet             & mission      & date              & $V$           & $r_e$ & $Q$    &  Ref. \\
                  &              &                   & [km s$^{-1}$] & [km]  & [$10^{28}$s$^{-1}$] & \\
\hline
1P/Halley         & VEGA, Giotto & March 1986        & 68--79        & 5.5   & 200.   & \citep{sagd+:1986, rein:1986} \\ 
19P/Borrelly      & Deep Space 1 & 22 September 2001 & 17            & 2.2   & 3.     & \citep{sode+:2002} \\
81P/Wild~2        & Stardust     & 2 January 2004    &  6            & 2.1   & 1.     & \citep{ahea+:2005} \\
9P/Tempel~1       & Deep Impact  & 4 July 2005       & 11            & 3.0   & 1.     & \citep{brow+:2004, brow+:2006}\\
67P/Churyumov-G.  & Rosetta      & 2014--2015        &  0            & 1.0   & 0.--1. & \citep{glas+:2006} \\
\hline
\end{tabular}
\end{center}

$V$ = flyby velocity.\\
$r_e$ = nucleus equivalent radius.\\
$Q$ = water production rate at time of exploration.\\
\index{comet!1P/Halley}
\index{comet!9P/Tempel 1}
\index{comet!19P/Borrelly}
\index{comet!81P/Wild 2}
\index{comet!67P/Churyumov-Gerasimenko}
\index{Giotto}
\index{Deep Space 1}
\index{Stardust}
\index{Deep Impact}
\index{Rosetta}
\end{table}

Jupiter-family and Oort-cloud comets are far from being equally well observed.

To show this for Earth-based observations, we will use the \textit{figure of
merit} $FM = Q$[H$_2$O]/$\Delta$, as introduced in \citep{mumm+:2002}.  This
parameter is roughly proportional to the expected signal, and allows us to
evaluate and compare the observability of comets.  Table~\ref{table:remote}
gives the figure of merit for Earth-based observations of recent long-period
comets, as well as for recent and future returns of short-period comets.  One
can see that unexpected, long-period comets offered much better opportunities
than short-period comets.  Indeed, the two best comets in the last twenty years
were, by far, C/1996 B2 (Hyakutake) and C/1995 O1 (Hale-Bopp).  Unprecedented
spectroscopic observations leaded to the detection of many new molecules in
these two comets.

\sloppy Short-period, Jupiter-family comets are all weak comets with $Q$
[H$_2$O] $\approx$ a few 10$^{28}$~s$^{-1}$ at most.  The best observing
conditions occur for these comets which make a close approach to the Earth.
This was recently the case for 73P/Schwassmann-Wachmann 3 ($\Delta = 0.08$ AU on
May 2006) and it will also happen for 103P/Hartley~2 in the future ($\Delta =
0.12$ AU on October 2010).  But the figure of merit of such Jupiter-family
comets is still much lower than that of the best long-period comets.

The situation is opposite with regard to space exploration
(Table~\ref{table:space}): short-period comets, which have expected returns, are
the only practicable targets for space missions, which are not yet versatile
enough to accommodate unexpected comets.  Flybys at a slow velocity and
rendezvous are only possible for ecliptic comets, due to the energy limitations
of current space technology.  1P/Halley is the only explored comet which does
not belong to the Jupiter family: the price to pay was a very large flyby
velocity ($\approx 70$ km~s$^{-1}$).  Although several studies of space missions
towards an unexpected comet were made \citep{pero+:1996}, there is currently no
firm plan for such a mission.

On the other hand, short-period comets have very well known orbits, allowing us
to perform accurate modelling of non-gravitational forces.  Asteroid-like
studies of cometary nuclei (i.e., measurement of albedo, size and light curve)
are also better done on these weak objects.

\section{Chemical diversity}
\label{section:chemistry}

Only very few comets were explored in situ, so the chemical composition of
comets and its diversity are basically investigated by remote sensing using
spectroscopy.  This investigation is indirect, since only the gas and dust coma
is observed, after the nucleus ices have sublimated.  Sublimation fractionation
may occur (i.e., the most volatile species sublimate first).  The production
rates of the sublimated molecules (or of their photodissociation products, the
so-called \textit{daughter molecules}) are determined.  How they are related to
the initial molecular abundances in the nucleus is an open issue.  Indeed, the
relative production rates vary as a function of heliocentric distances, as was
found for comet Hale-Bopp which was observed over a large range of heliocentric
distances \citep{bive+:2002-moni}.

Relative molecular production rates --- improperly named \textit{abundances} ---
are usually given relative to the water production rate, for heliocentric
distances $r_h \approx 1$ AU. At such distances, the outgassing of cometary ices
is dominated by water sublimation and it is likely that in this regime of strong
sublimation, all species more volatile than water are expelled without
significant fractionation.

\index{cometary ices} \index{parent molecules} \index{daughter 
molecules}

\begin{table}[ht]
\caption{Relative production rates of radicals.}
\protect\label{tab:radical-abundances}
\begin{center}
\begin{tabular}{lcccc}
\hline
 & \multicolumn{2}{c}{``typical"} & \multicolumn{2}{c}{``C$_2$-depleted"} \\
 & \multicolumn{2}{c}{C$_2$/CN $> 0.66$} & \multicolumn{2}{c}{C$_2$/CN $< 0.66$} \\
\hline
OH    & 100   &                 & 100    &                 \\
CN    & 0.32  &   (0.15--0.68)  & 0.20   &   (0.11--0.31)  \\
C$_2$ & 0.36  &   (0.13--0.79)  & 0.050  &  (0.007--0.10)  \\
C$_3$ & 0.025 & (0.0055--0.081) & 0.0066 & (0.0026--0.020) \\
NH    & 0.42  &   (0.17--1.6)   & 0.33   &   (0.11--1.2)   \\
\hline
\end{tabular}
\index{radical!CN}
\index{radical!OH}
\index{radical!NH}
\index{radical!C$_2$}
\index{radical!C$_3$}
\index{comets!typical comets}
\index{comets!carbon-depleted comets}
\end{center}

Adapted from \citep{ahea+:1995} where ``typical" and ``C$_2$--depleted" comets
are separated according to their C$_2$/CN ratio.  Mean production rates and
ranges (between parentheses) are listed relative to OH (normalized to 100).
\end{table}

\subsection{From the visible and the UV: daughter species.}

From narrow-band photometric observations of cometary radical, a database of
more than one hundred comets has now been constructed, with determinations of
the relative production rates of OH, CN, C$_2$, C$_3$, NH, NH$_2$
\citep{ahea+:1995, fink-hick:1996, fink:2006}.

\index{daughter molecules} \index{radical!CN} \index{radical!OH}
\index{radical!NH} \index{radical!NH$_2$} \index{radical!C$_2$}
\index{radical!C$_3$}

\begin{figure}
\centering
\includegraphics[width=6cm]{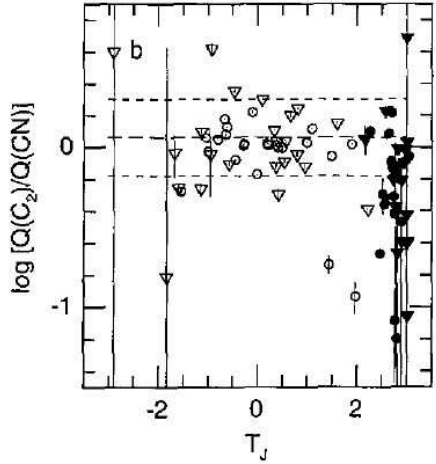}
\caption{The possible correlation between carbon-depleted comets and
Jupiter-family comets (filled symbols, with $T_J > 2$). 
From \citep{ahea+:1995}.}
\label{fig:depleted}
\index{comets!Jupiter-family comets}
\index{comets!carbon-depleted comets}
\index{Tisserand parameter}
\end{figure}

A'Hearn et al.\ \citep{ahea+:1995} proposed the existence of two classes of
comets, depending on their C$_2$/CN ratio, that they named ``typical" and
``C$_2$-depleted" comets.  From their statistics, about one-half of the
Jupiter-family comets are C$_2$-depleted, but this fraction is much smaller
among nearly-isotropic comets (Table~\ref{tab:radical-abundances} and
Fig.~\ref{fig:depleted}).

\begin{figure}
\centering
\includegraphics[width=10cm]{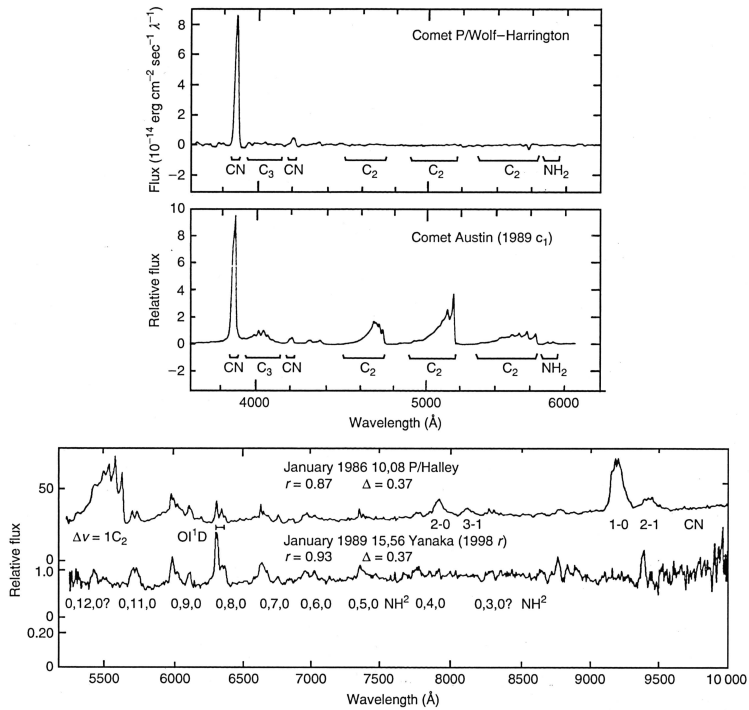}
\caption{Spectra of the \textit{anomalous} comets 43P/Wolf-Harrington
(C$_2$-depleted) and C/1988 Y1 (Yanaka) (NH$_2$-rich), compared to the
\textit{normal} comets C/1989 X1 (Austin) and 1P/Halley.  Adapted from
\citep{schl+:1993, fink:1992}.}
\label{fig:yanaka}
\index{comet!43P/Wolf-Harrington}
\index{comet!C/1988 Y1 (Yanaka)}
\index{comet!C/1989 X1 (Austin)}
\index{comet!1P/Halley}
\end{figure}

There are also occasionally comets with strongly anomalous compositions, which
are really puzzling monsters.  For instance, comet 43P/Wolf-Harrington showed
CN, but C$_2$ was not observed \citep{schl+:1993} (Fig.~\ref{fig:yanaka}); comet
C/1988 Y1 (Yanaka) showed NH$_2$, but no CN nor C$_2$ \citep{fink:1992}
(Fig.~\ref{fig:yanaka}).

Fink \citep{fink:2006} suggested a taxonomy similar to \citep{ahea+:1995}, but
with additional classes: (i) ``normal" comets (prototype: 1P/Halley); (ii)
comets with low C$_2$ and normal NH$_2$ (prototype: 9P/Tempel 1); (iii) comets
with low C$_2$ and low NH$_2$ (prototype: 21P/Giacobini-Zinner); and (iv) comet
C/1988 Y1 (Yanaka) which does not fit in the preceding classes.  The first three
classes comprise respectively 70, 22 and 6 \% of the comets in the database.

Unfortunately, these daughter-species studies do not inform us directly on the
nature of their parent species.  There is little doubt that the OH radical comes
mostly form the photolysis of water and indeed, the OH production rates are used
as a proxy for the cometary water production rates.  But there is no complete
understanding on which are the parents for the C$_2$, C$_3$ \citep{helb+:2005},
and even the CN radicals \citep{fray+:2005}.  There are also clues that some of
these species come, at least partly, from distributed sources (e.g., from
cometary grains), rather than from the nucleus ices.

\begin{table}
\caption{The composition of cometary volatiles from IR
observations$^{a)}$.}
\label{table:volatile_IR}
%\centering
\begin{tabular}{lccccccc}
\hline
comet                                    & CO      & CH$_4$   & C$_2$H$_6$ & C$_2$H$_2$ & HCN      & CH$_3$OH & Ref. \\
\hline
5 Oort-cloud comets$^{b)}$ & 1.8--17 & 0.5--1.5 & $\approx 0.6$ & 0.2--0.3 & 0.2--0.3 & $\approx 2$ & \citep{mumm+:2003} \\
1999 S4 (LINEAR)           & 0.9     & 0.2      & 0.1        & $< 0.12$   & 0.1      & $< 0.1$  & \citep{mumm+:2001} \\
C/2001 A2 (LINEAR)         & 3       & 1.4      & 1.6        & 0.4        & 0.5      & 3.5      & \citep{mage+:2007} \\ 
\hline
21P/Giacobini-Zinner       & $< 3$   &          & $< 0.08$   & $< 0.4$    & $< 0.3$  & 1.       & \citep{weav+:1999} \\
21P/Giacobini-Zinner       & 10      &          & 0.2        &            &          &          & \citep{mumm+:2000} \\
9P/Tempel~1 before DI      &         &          & 0.2        &            & 0.2      & 1.1      & \citep{mumm+:2005} \\
9P/Tempel~1 after DI       & 4       & 0.5      & 0.3        & 0.1        & 0.2      & 1.0      & \citep{mumm+:2005} \\
73P/S.-W.~3--C             &         &          & 0.2        & 0.2        & 0.3      &          & \citep{vill+:2006} \\
73P/S.-W.~3--B             & $< 2$   &          & 0.14       & $\approx 0.03$ & 0.25 & 0.2      & \citep{dell+:2006} \\
\hline
\end{tabular}

$^{a)}$ Production rates are given relative to water = 100.\\
$^{b)}$ C/1999 H1 (Lee), C/1995 O1 (Hale-Bopp), C/1996 B2 (Hyakutake),
153P/2002 C1 (Ikeya-Zhang), C/1999 T1 (McNaught-Hartley). 
\index{molecule!H$_2$O}
\index{molecule!HCN}
\index{molecule!CO}
\index{molecule!CH$_3$OH}
\index{molecule!CH$_4$}
\index{molecule!C$_2$H$_2$}
\index{molecule!C$_2$H$_6$}
\index{comets!Oort-cloud comets}
\index{comet!C/1999 S4 (LINEAR)}
\index{comet!C/2001 A2 (LINEAR)}
\index{comet!21P/Giacobini-Zinner}
\index{comet!9P/Tempel 1}
\index{comet!73P/Schwassmann-Wachmann 3}
\index{comet!C/1995 O1 (Hale-Bopp)}
\index{comet!C/1996 B2 (Hyakutake)}
\index{comet!C/1999 H1 (Lee)}
\index{comet!153P/2002 C1 (Ikeya-Zhang)}
\index{comet!C/1999 T1 (McNaught-Hartley)}
\end{table}

\begin{table}
\caption{The composition of cometary volatiles from radio
observations$^{a)}$.}
\label{table:volatile_radio}
\begin{tabular}{lcccccccl}
\hline
comet                  & CO          &  HCN        &  H$_2$S    & CS          & CH$_3$OH      & H$_2$CO    & Ref. \\
\hline
18 Oort-cloud comets  & $< 1.7$---23 & 0.08---0.25 & 0.12---1.5 & 0.05---0.17 & $< 0.9$---6.2 & 0.4---1.3  & \citep{bive+:2002-div} \\
\hline
22P/Kopff             & $< 6$        & 0.13        & 0.58       & $< 0.15$    & 2.5           &            & \citep{bive+:2002-div, bive:1997} \\
21P/Giacobini-Zinner  & $< 2.4$      & 0.09        & $< 0.36$   & 0.08        & 1.6           & 0.13       & \citep{bive+:2002-div} \\
19P/Borrelly (1994)   &              & 0.11        &            &             & 1.7           & 0.4        & \citep{bock+:2004} \\
19P/Borrelly (2001)   & $< 15$       & 0.06        & $< 0.45$   & 0.07        &               & $< 0.4$    & \citep{bock+:2004} \\
2P/Encke              &              & 0.09        & $< 0.4$    & $< 0.06$    & 4.1           & $< 1.4$    & \citep{bive+:2005} \\
9P/Tempel~1 before DI & $< 10$       & 0.11        & 0.5        & $< 0.13$    & 2.8           & $< 1.5$    & \citep{bive+:2007-Icarus} \\
9P/Tempel~1 after DI  & $< 31$       & 0.13        &            & $< 0.27$    & 2.7           & $< 2.3$    & \citep{bive+:2007-Icarus} \\
73P/S.-W.~3           & $<  6$       & 0.20        & 0.3        & 0.1         & 0.9           & 0.7        & \citep{bive+:2006-DPS} \\
\hline
\end{tabular}

$^{a)}$ Production rates are given relative to water = 100.\\
\index{molecule!H$_2$O}
\index{molecule!HCN}
\index{molecule!CO}
\index{molecule!CH$_3$OH}
\index{molecule!H$_2$CO}
\index{molecule!H$_2$S}
\index{radical!CS}
\index{comets!Oort-cloud comets}
\index{comet!21P/Giacobini-Zinner}
\index{comet!9P/Tempel 1}
\index{comet!73P/Schwassmann-Wachmann 3}
\index{comet!2P/Encke}
\index{comet!22P/Kopff}
\index{comet!19P/Borrelly}
\end{table}

\subsection{From the infrared and the radio: parent volatiles.}

Except for carbon monoxide which has strong electronic bands in the UV, parent
molecules have to be observed through the fluorescence of their vibrational
bands in the infrared, or their rotational lines at radio wavelengths
\citep{bock+:2005}.

About a dozen comets have been investigated by infrared spectroscopy
\citep{mumm+:2003, vill+:2006}.  A summary of the results is given in
Table~\ref{table:volatile_IR}.  The best studied Jupiter-family comet is
73P/Schwassmann-Wachmann 3 \citep{vill+:2006,dell+:2006}.

In addition to molecules such as H$_2$O, HCN, CH$_3$OH, H$_2$CO, NH$_3$ or OCS,
infrared spectroscopy can specifically study non-polar molecules, such as
hydrocarbons (CH$_4$, C$_2$H$_2$, C$_2$H$_6$), which do not have permitted
rotational lines.  It can also observe CO$_2$, but only from space due to strong
telluric absorption; thus this important cometary volatile has only been
observed in a very small number of comets.

\index{parent molecules} \index{molecule!H$_2$O} \index{molecule!HCN}
\index{molecule!CO} \index{molecule!CH$_3$OH} \index{molecule!CH$_4$}
\index{molecule!C$_2$H$_2$} \index{molecule!C$_2$H$_6$}
\index{molecule!NH$_3$} \index{molecule!OCS} \index{molecule!CO$_2$}

\begin{figure}
\centering
\includegraphics[width=0.8\hsize]{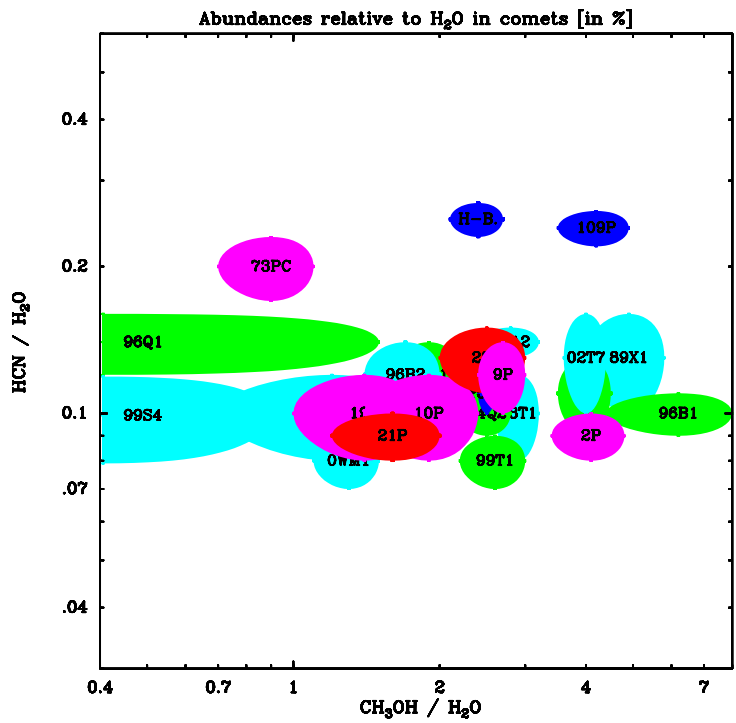}
\caption{The HCN and CH$_3$OH abundances relative to water from the
radio spectroscopic database of comets.  The ellipse sizes represent
errors.  From \citep{crov+:2007}.}
\label{fig:HCN-metha}    
\index{molecule!HCN}
\index{molecule!CH$_3$OH}
\end{figure}

Radio spectroscopy achieved detection of about 25 cometary molecules, radicals
and ions.  Fairly complex molecules, even with small abundances, can be studied
provided they have a significant dipolar moment.  Molecules as complex as methyl
formate (HCOOCH$_3$) or ethylene glycol ((CH$_2$OH)$_2$) have been identified in
comet Hale-Bopp.  About 30 comets have been investigated by this technique, with
resulting information on the production rates of H$_2$O, HCN, HNC, CH$_3$CN,
CH$_3$OH, H$_2$CO, CO, CS and H$_2$S. A preliminary analysis of the
comet-to-comet chemical diversity is given in \citep{bive+:2002-div} with
updates in \citep{bock+:2005, bive+:2006-AA, bive+:2007-Icarus}.  The summary of
the results is presented in Table~\ref{table:volatile_radio}.  Here again, the
best studied Jupiter-family comet is 73P/Schwassmann-Wachmann 3
\citep{bive+:2006-DPS}.  Whereas the HCN/H$_2$O ratio is relative constant,
several species, such as CH$_3$OH, H$_2$CO and H$_2$S, have important variations
from comet to comet .  As an example, the distributions of the HCN/H$_2$O and
CH$_3$OH/H$_2$O abundance ratios are shown in Fig.~\ref{fig:HCN-metha}.

\index{molecule!H$_2$O} \index{molecule!HCN} \index{molecule!HNC}
\index{molecule!CO} \index{molecule!CH$_3$OH}
\index{molecule!CH$_3$CN} \index{molecule!H$_2$CO}
\index{molecule!H$_2$S} \index{radical!CS} \index{molecule!HCOOCH$_3$}
\index{molecule!(CH$_2$OH)$_2$}

An extreme case is that of CO. In distant comets, such as C/1995 O1 (Hale-Bopp)
or 73P/Schwassmann-Wachmann 1, CO is the only observed cometary volatile.  At
$\approx 1$ AU from the Sun, the production rate of carbon monoxide relative to
water ranges from less than 1 \% (as observed in the UV by FUSE
\citep{feld+:2002}) to $\approx 30$ \% (C/1995 O1 (Hale-Bopp)).  The very origin
of cometary CO is a debated topic.  In situ observations of 1P/Halley and
infrared observations of C/1995 O1 (Hale-Bopp) pointed to a distributed source
in addition to the nucleus native source \citep{disa+:2001}, whereas radio
interferometric observations of C/1995 O1 (Hale-Bopp) are consistent with a
native source only \cite{bock+:2005-DPS}.

Two comets were remarked to be strongly ``volatile-depleted": C/1999 S4 (LINEAR)
\citep{mumm+:2001, bock+:2001} and 73P/Schwassmann-Wachmann 3 \citep{vill+:2006,
bive+:2006-DPS, dell+:2006}; Both have very low CO and CH$_3$OH abundances, but
normal HCN. We have here an example of a nearly-isotropic and a Jupiter-family
comet sharing the same anomalies!  (Indeed, these two comets experienced
fragmentation.)

The presence -- or not -- of hypervolatiles in cometary ices could be a
stringent clue to the formation temperature of comets.  CO, CH$_4$, N$_2$
condense at temperatures lower than $\approx 30$~K (Fig.~\ref{fig:sublime}).
The presence of N$_2$, tentatively inferred from the observation of N$_2^+$
bands, could not be confirmed in recent, high-quality observations
\citep{coch:2002}.  Noble gases (He, Ne, Ar, Kr\ldots) are still to be detected.
However, no clear difference in the abundance of very volatile molecules such as
CO, CH$_4$, C$_2$H$_2$, H$_2$S could be found between Jupiter-family and
Oort-cloud comets (Tables \ref{table:volatile_IR} and
\ref{table:volatile_radio}).

\index{hypervolatile} \index{noble gases} \index{molecule!CO}
\index{molecule!CO$_2$} \index{molecule!CH$_4$} \index{molecule!N$_2$}

\section{Diversity from physical indicators and others}
\label{section:physics}

\subsection{Spin temperatures.}

Molecules such as H$_2$O, NH$_3$, CH$_4$\ldots~which have several identical
hydrogen atoms exist in different spin species (\textit{ortho}--\textit{para},
$A$--$E$\ldots).  Spin transitions are forbidden, so that spin temperatures
could be preserved for a long time.  The ortho-to-para ratios (OPR) and spin
temperatures observed for water or for other species might thus be primordial.
First remarks on this topic were made in \cite{crov:1984, mumm+:1987}.

\index{ortho-to-para ratio (OPR)} \index{spin temperature}
\index{molecule!H$_2$O} \index{molecule!NH$_3$} \index{molecule!CH$_4$}

\begin{table}
\caption{Spin temperatures observed in comets.  Adapted from the
compilation of \citep{kawa+:2004}, and updated with recent results.} 
\center
\begin{tabular}{lcccc}
\hline
Comet                  & H$_2$O          & NH$_3$          & CH$_4$         & orbital period \\
		       & [K]             & [K]             & [K]            & [yr]   \\
\hline
1P/Halley              & $29\pm2$        &                 &                &      76 \\
C/1986 P1 (Wilson)     & $> 50$          &                 &                & dynamically new \\
C/1995 O1 (Hale-Bopp)  & $28\pm2$        & $26^{+10}_{-4}$ &                &  4\,000 \\
103P/Hartley 2         & $34\pm3$        &                 &                &       6.4 \\
C/1999 H1 (Lee)        & $30^{+15}_{-6}$ &                 &                & dynamically new \\
C/1999 S4 (LINEAR)     & $\geq 30$       & $27^{+3}_{-2}$  &                & dynamically new \\
C/2001 A2 (LINEAR)     & $23^{+4}_{-3}$  & $25^{+1}_{-2}$  &                & 40\,000 \\
C/2000 WM$_1$ (LINEAR) &                 & $30^{+5}_{-3}$  &                & dynamically new \\
153P/Ikeya-Zhang       &                 & $32^{+5}_{-4}$  &                &     365 \\
2P/Encke               &                 & $\ge 33$        &                &       3.3 \\
C/2001 Q4 (NEAT)       & $31^{+11}_{-5}$ & $31^{+4}_{-2}$  & $33^{+2}_{-1}$ & dynamically new \\
9P/Tempel 1            &                 & $24^{+2}_{-1}$  &                &       5.5 \\
C/2004 Q2 (Machholz)   & $> 35$          &                 &                & dynamically new \\
73P/Schwassmann-Wachmann 3 & $> 42$      &                 &                &       5.4 \\
\hline
\end{tabular}
\label{table:spin_temp}
\index{spin temperature}
\index{molecule!H$_2$O}
\index{molecule!NH$_3$}
\index{molecule!CH$_4$}
\index{ortho-to-para ratio (OPR)}
\index{comet!1P/Halley}
\index{comet!C/1986 P1 (Wilson)}
\index{comet!C/1995 O1 (Hale-Bopp)}
\index{comet!103P/Hartley 2}
\index{comet!C/1999 H1 (Lee)}
\index{comet!C/1999 S4 (LINEAR)}
\index{comet!C/2001 A2 (LINEAR)}
\index{comet!C/2000 WM$_1$ (LINEAR)}
\index{comet!153P/2002 C1 (Ikeya-Zhang)}
\index{comet!2P/Encke}
\index{comet!C/2001 Q4 (NEAT)}
\index{comet!9P/Tempel 1}
\index{comet!C/2004 Q2 (Machholz)}
\index{comet!73P/Schwassmann-Wachmann 3}
\end{table}

First determinations of the water OPR were made from air-borne infrared
observations of comets 1P/Halley and C/1986 P1 (Wilson) \cite{mumm+:1993}.  Then
accurate measurements were obtained with the \textit{Infrared Space Observatory}
on C/1995 O1 (Hale-Bopp) and Jupiter-family comet 103P/Hartley 2
\cite{crov+:1999,crov+:1997}.  Further results on water were obtained by
observing vibrational hot bands of water from the ground on bright comets.
These results were extended to ammonia.  The OPR of NH$_3$ itself cannot (yet)
be directly observed, but is derived from the OPR of NH$_2$ determined from its
visible spectrum \cite{kawa+:2001}.  The spin temperature of methane can also be
determined from the $E$, $A$, $F$ spin species relative populations measured
from its infrared spectrum.  \index{radical!NH$_2$} \index{Infrared Space
Observatory (ISO)}

All these results, recently reviewed in \cite{kawa+:2004} and summarized in
Table~\ref{table:spin_temp}, are quite puzzling: the observed spin temperatures
are remarkably clustered around 30~K, whatever the molecule, the heliocentric
distance of the comet or its dynamical history.  What is the signification of
this temperature?  It seems that any possible explanation can be ruled out:

\begin{itemize}

\item Equilibration within the coma would lead to spin temperatures depending on
the heliocentric distance, as is observed for the rotational temperatures of
cometary molecules.  Similar spin temperatures were observed for comets at $r_h
\approx 1$~AU and $r_h \approx 2.9$~AU. Laboratory experiments confirm that spin
temperatures are preserved in molecular jets.

\item Equilibration at the comet surface would also lead to spin temperatures
depending on the heliocentric distance, since surface temperatures vary roughly
as $r_h^{-1/2}$.  If equilibration occurs at sublimation, one would rather
expect $T_\mathrm{spin} \approx$ 150--200~K, the equilibrium temperature of
sublimating water ice in cometary conditions.  However, it would be interesting
to investigate in the laboratory whether the OPR is preserved during phase
transitions.

\item A spin temperature in equilibrium with the internal temperature of the
nucleus would nicely explain why the spin temperatures are the same for
different molecules.  However, the comet nucleus internal temperatures depend
upon the comet orbital history and are expected to differ between short-period
and long-period comets, whereas both classes of comets show the same
$T_\mathrm{spin}$.  On the other hand, the spin temperatures listed in
Table~\ref{table:spin_temp} pertain to the gas phase; the rotational levels of
molecular species have different energies in the solid phase, which leads to a
different correspondence between the OPR and $T_\mathrm{spin}$
\citep{limb+:2006}.

\item Although inter-spin conversions are forbidden, preservation of the spin
state over cosmological times seems to be highly unlikely \citep{jens+:2003,
limb+:2006}.  This is unfortunately difficult to test in the laboratory!  If
indeed the present spin temperatures reflect the temperatures at the formation
or condensation of the molecules this would imply that all comets formed in very
similar physical conditions.  Note also that H$_2$O, NH$_3$ and CH$_4$ have very
different condensation temperatures: equilibrium at condensation would lead to
different $T_\mathrm{spin}$ for these different molecules.

\end{itemize}

\begin{figure}
\centering
\includegraphics[width=10cm]{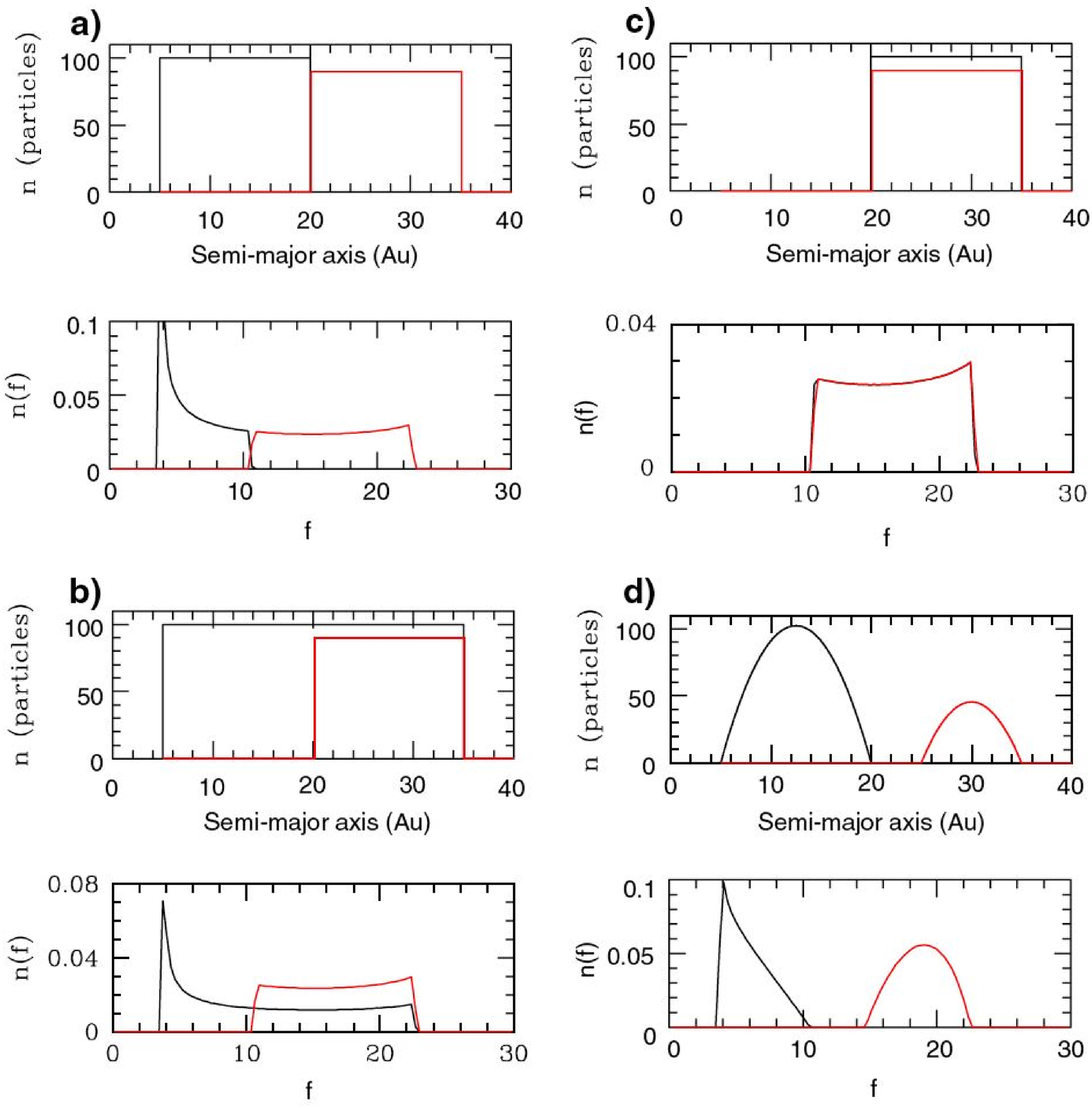}
\caption{Panels a) to d) show different examples of how the formation
regions of comets affect the distributions of their observed D/H
enrichment \textit{f}.  In each of the four panels, the upper diagram
shows the used population distributions, and the lower diagram the
\textit{f} distribution that would result.  The red line shows
Kuiper-belt objects, and the black Oort-cloud objects.  From
\citep{horn+:2006}.}
\label{fig:mousis}
\index{comets!Oort-cloud comets}
\index{Kuiper-belt objects}
\index{isotopic species}
\end{figure}

\begin{table}
\caption{Nitrogen and carbon isotopic ratios in comets from
observations of the CN radical.  Adapted from \citep{huts+:2005} and
\citep{jehi+:2006}.}
\label{table:isotopic_NC}
\centering
\begin{tabular}{llcc}
\hline
comet                         & $^{14}$N/$^{15}$N & $^{12}$C/$^{13}$C \\
\hline
C/1995 O1 (Hale-Bopp)         & $140 \pm 35$ &  $90 \pm 30$ \\
C/2000 WM$_1$ (LINEAR)        & $140 \pm 30$ & $115 \pm 20$ \\
122P/de Vico                  & $140 \pm 20$ &  $90 \pm 10$ \\
153P/2002 C1 (Ikeya-Zhang)    & $170 \pm 50$ &  $90 \pm 25$ \\
C/2001 Q4 (NEAT)              & $135 \pm 20$ &  $90 \pm 15$ \\
C/2003 K4 (LINEAR)            & $135 \pm 20$ &  $90 \pm 15$ \\
C/1999 S4 (LINEAR)            & $150 \pm 40$ & $100 \pm 30$ \\
88P/Howell                    & $140 \pm 15$ &  $90 \pm 10$ \\
9P/Tempel~1                   & $145 \pm 20$ &  $95 \pm 15$ \\
\hline
\end{tabular}
\index{radical!CN}
\index{isotopic species}
\index{comet!C/1995 O1 (Hale-Bopp)}
\index{comet!C/2000 WM$_1$ (LINEAR)}
\index{comet!122P/de Vico}
\index{comet!153P/2002 C1 (Ikeya-Zhang)}
\index{comet!C/1999 S4 (LINEAR)}
\index{comet!C/2001 Q4 (NEAT)}
\index{comet!C/2003 K4 (LINEAR)}
\index{comet!9P/Tempel 1}
\index{comet!88P/Howell}
\end{table}

\subsection{Isotopic ratios.}

Isotopic ratios are a crucial diagnostic of the physico-chemical conditions of
the early Solar Nebula.  This is discussed in detail in \citep{altw-bock:2003,
bock+:2005} and the observational results are summarized in Table~3 of
\citep{bock+:2005}.

The only comets for which the D/H ratio in water has been firmly determined ---
1P/Halley, C/1996 B2 (Hyakutake) and C/1995 O1 (Hale-Bopp) --- are three
Oort-cloud comets.  All three determinations are close to D/H $\approx 3 \times
10^{-4}$ (i.e., twice the terrestrial value and a factor of ten larger than the
protosolar value).  It would be crucial to determine D/H for Jupiter-family
comets.  But the observation of HDO, whose main rotational transitions are in
the submillimetric domain, is difficult.  \index{molecule!HDO}

Some chemical models of the primitive Solar Nebula predict a strong dependence
of the D/H ratio on the heliocentric distance $r_h$.  So, it is in principle
possible, from the distribution of the D/H enrichment of a class of comets, to
determine the $r_h$ distribution of its formation site.  (See \citep{horn+:2006}
and Fig.~\ref{fig:mousis}).

Puzzling results were obtained for the $^{14}$N/$^{15}$N isotopic ratio.  From
high-resolution visible spectra of the CN radical in several comets,
$^{14}$N/$^{15}$N $\approx 150$ was consistently observed
(Table~\ref{table:isotopic_NC}, \citep{huts+:2005}), whereas $^{14}$N/$^{15}$N
was 300 (close to the terrestrial value) from a radio line of HCN observed in
comet Hale-Bopp \citep{jewi+:1997}.  In contrast, the $^{12}$C/$^{13}$C ratio is
found to be $90 \pm 4$ in CN for the whole sample, close to the terrestrial
ratio (Table~\ref{table:isotopic_NC}, \citep{huts+:2005}).  This points to an
additional source of CN, other than HCN and heavily enriched in $^{15}$N, which
is still to be identified.  High molecular weight organics such as polymerized
cyanopolyynes were invoked (see \citep{fray+:2005} for a discussion of the
sources of the CN radical).  But surprisingly, the $^{14}$N/$^{15}$N ratio does
not vary from comet to comet, whereas these objects had strongly different
dust-to-gas ratios.  This isotopic ratio is also exactly the same for
Jupiter-family and Oort-cloud comets.

\subsection{Dust properties.}

The presence of crystalline silicates in cometary dust is a touchstone for
models of cometary formation and Solar System history.  The existence in
cometary material of silicates in both crystalline and amorphous phases
testifies to the existence of radial mixing in the primitive Solar Nebula
\citep{bock+:2002}.  \index{crystalline silicates}

Crystalline silicates (forsterite) were clearly identified in the infrared
spectrum of comet C/1995 O1 (Hale-Bopp) observed by \textit{ISO}\index{Infrared
Space Observatory (ISO)} \citep{crov+:1997, crov+:2000}.  They were then
observed in several other Oort-cloud comets \citep{hann-brad:2005}.

The presence of crystalline silicates in Jupiter-family comets was more
difficult to assess.  They were finally identified in the infrared spectra of
103P/Hartley 2 \citep{crov+:2000}, 78P/Gehrels 2 \citep{oots+:2007}in 9P/Tempel
1 (\textit{Spitzer} observations after \textit{Deep Impact} \citep{liss+:2006})
and tentatively in 29P/Schwassmann-Wachmann 1 \citep{stan+:2004}.  They are also
present in the dust samples of 81P/Wild 2 brought back by \textit{Stardust}
\citep{brow+:2006, zole+:2006}.

\index{Deep Impact} \index{Stardust} \index{Spitzer Infrared Telescope}
\index{comet!103P/Hartley 2} \index{comet!9P/Tempel 1} \index{comet!78P/Gehrels
2} \index{comet!81P/Wild 2} \index{comet!29P/Schwassmann-Wachmann 1}

It seems that crystalline silicates are present in all classes of comets, but
that the silicate feature, relative to continuum, is (much) weaker in
Jupiter-family comets.  So, it was more difficult to identify crystalline
silicates in Jupiter-family comets just because these comets are more difficult
to observe.

Two different classes of comets are also distinguished from the polarimetric
properties of cometary dust \citep{leva+:1996, kolo+:2007}, apparently
correlated with the strength of the silicate feature and the period of the
comet.  It thus appears that these different dust properties are due to the
evolution of the nucleus surface properties and the building up of a mantle for
short-period comets.

\subsection{Nucleus properties}

Available data on comet nucleus sizes are reviewed in \citep{lamy+:2005}.
Clearly, a selection effect is at work here: whereas many short-period comets
were investigated, only the largest nuclei of long-period comets were observed.
Thus it is little wonder that the observed sizes are larger for long-period than
for short-period comets.

Direct data on comet nucleus density are lacking.  Most determinations are based
upon the modelling of non-gravitational forces \citep{weis+:2005}.  They are
notoriously unreliable due to the complexity of the physical processes at work.
Such modelling is almost exclusively available for short-period comets for which
several returns are documented.  Measurements from space exploration are only
expected for Jupiter-family comets (e.g., 67P/Churyumov-Gerasimenko with
\textit{Rosetta}).  A determination from the ballistics of the \textit{Deep
Impact} ejecta was attempted for 9P/Tempel 1 \citep{ahea+:2005}.  \index{Deep
Impact} \index{Rosetta}

Available data on comet nucleus albedos and colours are also reviewed in
\citep{lamy+:2005}.  These parameters are similar for nearly-isotropic and
ecliptic comets.  The range of nucleus albedos is remarkably narrow ($0.04 \pm
0.02$).

\section{Discussion and conclusions}
\label{section:discus}

In the preceding sections, it was shown that comets have indeed a large range of
dynamical and physico-chemical properties.  It was investigated whether these
diversities could be interpreted in the frame of a consistent formation
scenario.  The following points were assessed:

\begin{itemize}

\item Most comets can be unambiguously separated into two classes:
nearly-isotropic comets and ecliptic comets.  Although there might be some
interlopers and ill-classified objects (e.g., Halley-type comets with small
inclination), this classification seems to be rather robust.
    
\index{comets!nearly-isotropic comets} \index{comets!ecliptic comets}

\item These two classes are recognized to originate from two distinct
reservoirs.  Nearly-isotropic comets are coming from the Oort cloud.  Ecliptic
comets are coming from a nearer, disc-like trans-Neptunian reservoir which was
first identified with the Kuiper belt itself, but which is now believed to be
the scattered disc of trans-Neptunian objects.
    
\index{Oort cloud} \index{Kuiper belt} \index{scattered disc}

\item In none of these reservoirs, comets appear to have formed in situ.  They
were formed in the inner Solar System, and then expelled to the outer Solar
System (or outside the Solar System) through gravitational interaction with the
giant planets.  It was formerly believed that nearly-isotropic comets were
indeed formed in the Kuiper belt (hence the analogy between Jupiter-family
comets and Kuiper-belt comets), but this hypothesis is no longer tenable in view
of dynamical simulations.

\item The very sites of cometary formation are still ill-determined.  The place
of the giant planets --- their distance to the Sun and even their order --- have
evolved in the early stages of the Solar System, complicating the formation
scenario.
    
\item Our study of nearly-isotropic and ecliptic comets suffers from a strong
observational bias.  The former class comprises the brightest objects which
allow us to perform the best remote spectroscopic investigations.  In contrast,
the latter class is the preferred target for space exploration.
    
\item Narrow-band photometry provides us with the largest (in terms of number of
comets investigated) statistical investigation of the cometary chemical
composition.  A class of carbon-depleted (C$_2$--poor) comets has been clearly
identified and is possibly associated with Jupiter-family comets.  However, we
do not know how this translates into abundances of definite nucleus ice
molecules.
    
\index{comets!carbon-depleted comets}

\item Infrared and radio spectroscopy can identify specific ice molecules, but
the number of investigated comets is still limited (especially for
Jupiter-family comets).  A large diversity in parent-molecule relative
abundances has been observed, but no obvious correlation with dynamical classes
could be found.

\item There is no observed correlation between the abundance of hypervolatiles
and the dynamical class.  For instance, carbon monoxide is present with either
low or high abundance in both classes of comets.
    
\index{hypervolatile}

\item Other indicators, such as the ortho-to-para ratio or the $^{14}$N/$^{15}$N
isotopic ratio, which are believed to be linked to cometary origins, are
remarkably similar for different dynamical classes, pointing to uniform
formation conditions.  But we must point out how poor is our understanding of
the meaning of the cometary ortho-to-para ratio, and we are still awaiting for
an explanation to the different $^{14}$N/$^{15}$N ratios observed for CN and HCN
in comet Hale-Bopp.
    
\index{ortho-to-para ratio (OPR)} \index{isotopic species}
\index{comet!C/1995 O1 (Hale-Bopp)}

\item The observations of split (or progressively fragmenting) comets show no
sign of nucleus heterogeneity, suggesting that these objects accreted from
planetesimals formed in uniform conditions.
    
\index{split comets} 
    
\item The role of aging (and of the so-called ``space weathering" process) is
uncertain.  One would expect short-period comets to experience significant
sublimation fractionation and a depletion of hypervolatiles in the outer layer
of their nucleus \citep{meec-svor:2005, pria+:2005}.  As noted above, no
depletion of the CO production is observed for short-period comets compared to
dynamically new comets.  No conclusive change in the chemical composition of the
material excavated by \textit{Deep Impact} on comet 9P/Tempel 1 was found, if
one excludes an enhancement of organic compounds in the infrared spectra
\citep{ahea+:2005} (but in our opinion, this preliminary finding should await
confirmation by considering optical depth effects \citep{crov:2006}).
    
\index{Deep Impact} \index{comet!9P/Tempel 1}

\item Forgetting about dynamical classes, we could investigate whether relative
molecular abundances point to different formation regions in a consistent way.
Following Fig~\ref{fig:sublime}, we would expect correlations between the
abundance of molecular species and their sublimation temperatures, if comets
indeed formed from regions of significantly different temperatures.  No such
correlation could be yet found.  (For instance, from an investigation of 7
Oort-cloud comets, no correlation could be found between the hypervolatiles CO
and CH$_4$, whereas their abundances relative to water varied by factors 3--4
\citep{gibb+:2003}.)  Thus the possible existence of different sites of
formation is still an open question.

\end{itemize}

In conclusion, why comets are so different when some properties are considered
--- and why some other properties are remarkably uniform --- is not yet
understood.  Further investigations are certainly needed, especially to reliably
assess the composition of a large sample of comets.  Space exploration is
helpful for the deep study of specific, hopefully representative objects.  But if
we aim at a taxonomic approach, progress will come from the statistical
investigation by remote sensing of a large number of objects.
    
\bibliographystyle{naturemag}
\bibliography{Crovisier_Blois}

%\printindex

\bigskip
\bigskip

\noindent \textit{Based upon ``Diversity of comets: the two
reservoirs'', an invited review presented at the XVIII\`emes
Rencontres de Blois: Planetary Science: Challenges and Discoveries,
28th May -- 2nd June 2006, Blois, France.}

\bigskip

\noindent \today 

\end{document}